# On defining and modeling context-awareness

Panteleimon Rodis,
Hellenic Open University.
std123937@ac.eap.gr 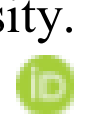

## Abstract

**Purpose -** This paper presents a methodology for defining and modeling context-awareness and describing efficiently the interactions between systems, applications and their context. Also the relation of modern context-aware systems with distributed computation is investigated.
**Design/methodology/approach -** On this purpose, definitions of context and context-awareness are developed based on the theory of computation and especially on a computational model for interactive computation which extends the classical Turing Machine model. The computational model proposed here, encloses interaction and networking capabilities for computational machines.
**Findings -** The definition of context presented here develop a mathematical framework for working with context. Also the modeling approach of distributed computing enables us to build robust, scalable and detailed models for systems and application with context-aware capabilities. Also enables us to map the procedures that support context-aware operations providing detailed descriptions about the interactions of applications with their context as well as with other external sources.
**Practical implications -** A case study of a cloud based context-aware application is examined using the modeling methodology described in the paper so as to demonstrate the practical usage of the theoretical framework that is presented
**Originality/value -** The originality on the framework presented here relies on the connection of context-awareness with the theory of computation and distributed computing.

## 1. Introduction

The modeling methodology presented in this paper aims to provide mathematical foundations in the discussion around context-awareness. In the rich literature on the subject, researchers usually use plain language on defining context-awareness which sometimes leads to vague or disputed definitions; see the discussion about weak or inadequate definitions in (Abowd et al., 1999). Moreover, the connection of context-awareness with other fields of computer science like computational complexity or distributed systems is not clearly defined. On these matters, the contribution of this paper is the development of a framework for the definition of context-awareness based on the theory of computation which provides an important insight to the subject. The presented framework is later used for the development of a robust and scalable modeling methodology for context-aware and distributed systems based on the theory of computation. The modeling methodology aims to provide tools for mapping the structure and functionality of context-aware systems taking into account the support for distributed computation of modern computational systems.
There are two more issues that must be investigated before presenting our framework.
The first is the user-centric perception about context-awareness that is widely spread in the literature. Adopting a user-centric, a data-driven or a system-centric approach in defining context and developing a modeling methodology determines the degree of generality that the produced models will have and how restrictive the designing assumptions will be.
The second issue under investigation is the relation between modern implementations of context-awareness and distributed computation. The increased use of cloud services and networked software components in the development of systems and applications gradually moves research about software design from stand alone systems and applications to distributed computational systems. This tension affects the technology concerning context-aware systems and as a result a modeling framework must consider distributed computation.
In the next section let us discuss in details the issues that the methodology on this paper addresses to. In section 3 a computational model appropriate for modeling interacting software systems is defined. In section 4 the computational model is used as a basis for providing a concrete definition of context-aware systems. In section 5 we develop a modeling methodology for systems of distributed computing with context-aware capabilities. A case study of a context-aware application is presented in section 6 demonstrating the modeling capabilities of the definition proposed followed by some discussion on section 7.

# 2. Issues on defining and modeling context

## 2.1 Context definitions

The research concerning context-awareness usually focuses on computational systems that detect the environment they operate in, which constitutes their context of operation, and use this information to provide services to their users.
A comprehensive definition of a widely accepted perception of context-awareness that summarizes a big part of the rich literature on the subject is proposed in (Abowd et al., 1999).

"A system is context-aware if it uses context to provide relevant information and/or services to the user, where relevancy depends on the user's task."

This definition although comprehensive and sufficient restricts context-awareness in systems that interact with users and have a user-centric operation. It does not take into account software components that use context-awareness in order to produce information for other systems or other software components besides users. Context under this consideration is restricted to be a part of the user's environment ignoring the cases where a system is a part of a network or it supports distributed computing operations and the context of the system is extended beyond the user's environment.
An interesting perception of context is presented in (Coulouris et al., 2012). The authors identify that the physical circumstance of a user or device may be relevant to the behavior of a system that may be mobile or even distributed. Context of an entity is then an aspect of its physical circumstances of relevance to system behavior. In this statement an entity may be a human, a place or a device and its contextual information includes relatively simple values that describe the environment of the entity as well as the state of other associated entities. Contextual information is perceived by sensors that detect the environment and other entities.
This perception of context is closer to modern computational systems. However, it does not determine the context of autonomous software applications that interact with other software components. A software application may be installed in some cloud infrastructure, operate autonomously and interact with other systems outside the cloud; for a survey covering this kind of awareness see (Truong, 2009). The context for this application is consisted of the systems that it interacts. Interactions of this kind are achieved by exchanging messages among the interacting systems, not by sensors, and one system is not necessarily aware of the state of the other systems that it interacts with.
Other works regard operational (Zimmermann, 2007) or conceptual (Dourishl, 2004) definitions of context. In each of them the authors define the use and perception of context by the properties or the operation of the entities that constitute an application and its context. These approaches suppress generality and restrict the discussion by focusing on special characteristics of the entities under investigation.
The above mentioned works reflect a wider approach of the literature on the subject which avoids the use of mathematical foundations in the discussion around the definition and the essence of context-awareness. Since its early days, computer science is based on mathematical foundations that provide a solid basis for developing theory and practice. It would be fruitful to follow this approach for the research around context-awareness as well; this is a goal of this paper.

## 2.2 On the user-centric approach

Influenced by the vision of pervasive computing the discussion around context and context awareness usually adopts a user-centric orientation. Context-awareness in this case is considered as a set of procedures oriented on detecting the users environment, so as to collect the necessary information which will then be used as input for operations that provide services to the users.
The issue that is ignored in this perception is that there are systems equipped with context-aware operations that operate without users; or at least the users do not play a central role in their operation. We may refer to autonomous systems and to artificial intelligence (AI). Let us argue on this perception by providing some examples.
If not already implemented, it is feasible to consider an AI system that controls the floodgates of a dam equipped with a hydroelectric power station. Such a system may take into account the weather conditions, the volume of water on the dam and the need for production of electric power in order to adjust the floodgates so as to avoid flooding and achieve an optimum performance for the station. This system is context-aware and the users are really absent.
A simple example is also the goal for autonomous self-driving cars. The goal is to produce cars that will serve the users automatically. Autonomous cars need to detect their environment, traffic, road and weather conditions in order to achieve the goal of efficiently transporting passengers to a predefined destination. In this example, if we replace passengers with luggage it becomes clear that the existence of users is not necessary.
Another common example includes software entities that operate in a computer network, where their environment and context is considered to be the network devices and the other software components like network services. In this case the software entities need to identify and monitor their artificial environment and then interact with it in order to operate, regardless if they eventually reach any users or not.
It is more useful to adopt a system-centric perception when discussing context-awareness. On these examples, it becomes clear that in many cases the users if not absent at least they do not play a critical role in the operation of context-aware systems.

### 2.3 Distributed systems instead of standalone applications

The discussion around modern system engineering should move beyond stand alone systems and applications and consider systems that support distributed computing. The extensive use of networks and cloud infrastructures in modern systems and the development of architectures like Platform as a Service (P.a.a.S) and Software as a Service (S.a.a.S), see (Dinh et al., 2013), leads to the development of distributed systems rather than stand alone applications. In the work of (Truong, 2009) the support of distributed computing is a demanding need for every context modeling approach. In a distributed system its components are hosted in different locations interacting with each other, with other systems and with context. On this basis, the framework presented in this paper regards systems that support distributed computing.

## 3. A model of interactive computation

In this section let us describe computational systems using a computational model, the model is valid for distributed and stand alone systems. The advantage in this approach is a solid definition for the system and its interaction with its context which then may be used in the analysis of system structure and operation. It is necessary to use computational models that support interactive computation; this choice is enforced by the interactive nature of the systems studied in this paper. Let us then build a computational model on Turing Machines enhanced with interactive and networking capabilities, called Networked Turing Machine. The ideas that structure this model derive from the choice machines of Alan Turing (Turing, 1936) and from the interactive computation models of Peter Wegner (Wegner, 1998) and Persistent Turing Machines (Goldin, 2000).

A Networked Turing Machine $m$ is a multi-tape Turing Machine consisting of a working tape and a number of input tapes. On the input tapes, $m$ reads symbols printed from other machines or other entities. The symbols on the input tapes may be printed before or during the operation of $m$. Reading and writing on an input tape $f$ is performed by different entities, two or more machines or other sources, so we have to develop a mechanism that will handle the operation of the tape properly.

Tape $f$ is equipped with two heads; head-1 prints symbols on the tape and it is control by the machines interacting with $m$; head-2 reads the symbols and it functions in compliance with the transition function of $m$. Both heads may scan the same cell without conflicting but they may not operate at the same time. Then $f$ operates under the following rules

1. Initially all the cells in $f$ have the blank symbol printed on them.
2. Both heads initially move to the first cell.
3. On input of symbol $s$ from machine $n$. Head-1 prints $s$ and moves to the right.
4. On request to read. If head-2 reads the blank symbol stands still, otherwise reads the symbol and moves to the right.

The working tape is used for performing computations as specified on its transition function considering the symbols read from the input tapes and the symbols on the working tape. Machine $m$ prints symbols on the input tapes of other Machines; this is the useful output of the computations performed on the working tape of $m$.

*Definition*. Networked Turing Machine $m$ that consists of $i$ input tapes, one working tape and prints output on $o$ input tapes of other machines is an 6-tuple ($Q$, $\Sigma$, $\Gamma$, $\delta$, $q_0$, $q_f$), where

1. Q is a finite set of states
2. $\Sigma$ is a finite input alphabet, which includes the blank symbol
3. $\Gamma$ is the tape and output finite alphabet $\Sigma \subseteq \Gamma$
4. $\delta: Q \times \Gamma \rightarrow Q \times (\Gamma \times \{L, R, S\}) \times (\Sigma^i \times \{R, S\}^i) \times (\Gamma^o \times \{R, S\}^o)$ the transition function
5. $q_0$ is the start state
6. $q_f$ is the halting state

In the transition function, R indicates that the head of a tape moves to the right, L to the left and S that the head stands still. For the working tape all choices are valid. The heads in the input tapes if not still move only to the right reading the symbols already printed. The heads printing output symbols also moves only to the right if not still.

As a computational model a Networked Turing Machine is as powerful as a Turing Machine.

*Equivalence theorem*. Let $m$ be a Networked Turing Machine and $r$ a one-tape Turing machine. Machine $r$ may simulate the operation of $m$ and $m$ may simulate the operation of $r$.

*Proof*. Machine $m$ runs the transition function of $r$ in its working tape and thus simulates the operation of $r$ the heads in its input tapes stand still and there is no output produced.

Let $h$ be a Turing machine that prints symbols on the input tapes of $m$. We build $l$-tape Turing machine $q$ that simulates the operations of both $m$ and $h$. In the first $i$ tapes of $q$ the transition function prints symbols according to the transition function of $h$. Tape $i$+1 is the working tape as in $m$. In the rest tapes the output of $m$ is printed. On this way $q$ simulates the operation of $m$ and its interaction with given machine $h$.

The operation of every multi-tape machine may be simulated by a one-tape machine and this stands for $q$ and $r$ as the transition function of the tapes of $m$ is clearly Turing computable.

The advantage in using Networked Turing Machines to model computational systems lays on the interaction capabilities that they enclose. Let us next define interaction and networking among Networked Turing Machines.

*Connectivity*. Machines $m$, $n$ are considered connected iff $m$ may print symbols on the input tapes of $n$, then predicate $con(m,n)$ is true.

*Networking*. Set $N$ forms a network iff
$\forall m \exists n (m \in N \wedge n \in N) \rightarrow con(m,n) \vee con(n,m)$.
Network $N$ interacts with external source $q$ when
$\exists m \exists q (m \in N \wedge q \notin N) \rightarrow con(m,q) \vee con(q,m)$.

Interaction among machines and the computations performed in a network need to have some rhythm. The machines that constitute a network simulate real world operations and real world operations require some time to get executed. Real world operations may run on hardware of different speeds. Some operations will run on faster hardware than others and this must be modeled in our network of Turing machines. So, the times $t_b$, $t_c$ that are required by machines $b$ and $c$ respectively in order to transit between two consecutive states in given moment $d$ must form ratio $\rho_d = t_b / t_c$.
For this, on network $N$ let us define fixed time step $t$ and for each machine $m$ let us define number $\sigma_m$, so that each transition between states as defined in the transition function of $m$ is executed in time $t_m = t / \sigma_m$. As a result for $b$ and $c$ in $d$ stands that $\rho_d = \sigma_c / \sigma_b$. On this way each machine has its own speed of operation and interaction with other machines. For the case study in section V we consider all machines working on the same speed.
The computational model presented so far may be used in modeling computational systems. Let $A$ be a computational system and $T$ a network of Networked Turing machines so that each computational operation $a_i$ of $A$ is simulated by a Networked Turing machine $i \in T$ and every machine in $T$ corresponds to an operation of $A$. Interaction between the distinct operations of $A$ is modeled as interaction between machines in $T$. The functionality of $A$ may then be modeled as a directed graph $G = (T, I)$ where the elements of $T$ are denoted by the set of nodes of $G$ and $I$ is the set of directed edges of the graph that denote the possible connections among the elements of $T$; for $con(i, j)$ there is directed edge in $G$ from node $i$ to $j$.
The networked machines that we used, model the computational operations of $A$ and not the physical devices in which they are carried. So, machine $A_i$ may model an operation carried in more than one computational devices of $A$ and any subsystem of $A$ may execute more than one operations modeled by more than one machine.
There is no real restriction on the level of detail that a system may be modeled using Networked Turing machines. Each procedure $a_i$ of $A$ may be analyzed in a number of sub-procedures $\{a_{i,1}, a_{i,2}, ...\}$ that their operations constitute $a_i$. Whether a single machine $m$ will be defined on $T$ so as to represent $a_i$ or a sub-network $T_i$ will be developed on $T$ so as to represent the sub-procedures $\{a_{i,1}, a_{i,2}, ...\}$ in details it is a matter of choice for the system designer. Describing software systems using networks of Turing machines provides the system designers with the flexibility of actually presenting on a model the level of details that they care to demonstrate about their system. For system $A$ we may build many networks where each network models $A$ in deferent level of details addressing different recipients. This feature enables a scalable approach in every system that is modeled.
Next, we define a framework for modeling the interaction between $A$ and its context.

# 4. Definition of context-awareness

Let $e$ be the environment in which computational system $A$ operates. On the term "environment" we consider the physical and artificial surroundings of $A$; the real, man-made or virtual content of the space in which $A$ is embodied.
We define set $E$ as the set of variables that describe the properties of $e$ in any given moment. For any variable $v_i \in E$ we consider vector $V_i = \{v_t, v_{t+1}, v_{t+2}....\}$ which stores the evaluations of $v_i$ in moments $t$, $t$+1, $t$+2 etc; where $t$ is a fixed time step. We then denote as $v_{i,t}$ the evaluation of variable $v_i$ in moment $t$. It is reasonable to restrict the variables in $E$ and the evaluations stored in each $V_i$ only to those that were valid during the period of the operation of $A$. As a consequence, in a changing environment $e$ the elements of $E$ change according to the changes in $e$ for any given moment. Set $E$ then should not be considered as a static factor in all systems but rather as a dynamic set that may be subject to changes over time.
The type of the variables in $E$ is not limited to numerical variables but it has a more general sense. Any variable $v_i$ may have any programmable accepted evaluation; any evaluation that could be accepted by a Turing machine. Moreover, any user interaction with $A$ is denoted by a set of vectors that store the parameters of the interaction in the time period of the interaction. Users are part of $e$ and their actions are described by some variables in a subset of $E$. The information that users enter in $A$ modifies the input of some $m \in T$.
Interaction of $A$ with $e$ is defined as the variation of some variables in $E$ as a cause of the operation of $A$. In this case, there is operation $a$ performed by $A$, simulated by Turing machine $A_a$, and there is vector $V_i$ in $E$ so that for fixed $t$ evaluation $v_{i,t}$ equals the output of $A_a$ in moment $t$. The information that $A$ displays to its users in time $t$ causes the variation of some variables of $E$.
Now let $C$ be the set of all vectors $V_i$ for environment $e$.

*Definition of Context-Awareness*. System $A$ is Context-Aware iff there is $C_A \subseteq C$ so that for each evaluation $v_{i,t}$ of vector $V_i \in C_A$ there is $A_i \in T$ that accepts $v_{i,t}$ and produces an output. Set $C_A$ describes the context of $A$.

As analyzed in chapter 7 of Posland's book about Ubiquitous Computing (Poslad, 2011) for some systems there is the notion of goal. A goal for system $A$ is to produce desired output $o$ on input of context parameters that are measured using context-awareness.
In many cases accessing the data that $A$ actually needs to produce a goal output is difficult or even impossible for a variety of reasons technological, social or any other. In these cases we are obliged to approximate the context of $A$ so as to achieve an effective approximation of our goal. On this basis let us next define the notion of effective Context-Awareness.

*Definition of effective Context-Awareness*. System $A$ is effectively Context-Aware when during its operation it accepts as input $C_A' \subset C_A$ and produces an output relatively similar to the output it produces when inputted $C_A$.

This definition provides a subjective criterion for effective Content-Awareness. It is in the judgment of the system designer to characterize the output of a system as "relatively similar" to the goal and certainly the criteria of doing so are case-dependent.

# 5. Modeling context-aware distributed systems

## 5.1 Modeling literature

In literature, modeling context-awareness is built in more solid ground than its definitions. Researchers use tools from software engineering like UML, ontology and mathematical logic (Baldauf et al., 2007) in order to develop efficient modeling methodologies for context-aware systems.
The use of UML and object-oriented methodologies in modeling context aims to provide a description of the information that applications derive from their environments and then how this information is used; see (Bardram, 2005), (Bettini, et al., 2010), (Bikakis et al., 2007), (Henricksen et al., 2003, 2006), (Kapitsaki et al., 2009), (Niu and Wang, 2016), (Sheng and Benatallah, 2005), (Strang and Linnhoff-Popien, 2004), (Vale, 2008), (Yu et al., 2010). These approaches target to building high level functional and conceptual descriptions of applications and the way they treat context.
Other modeling frameworks include logic (McCarthy, 1993), conceptual modeling languages (Hoyos et al., 2010, 2013) and ontology (Baumgartner, 2005), (Korpipaa et al., 2003). These methodologies provide conceptual approaches to what is considered to be context for every application and it is useful in the functionality of the application. These frameworks are designed to model the services provided by each system and the reason why they are built to function as they do.
The above frameworks, although valuable, do not relate context-aware operations with the actual structure of the system as they mostly focus on a description of the usability and functionality of the system and the services it provides. Also, most of them do not support modeling distributed systems.
The methodology presented next aims to supplement the above mentioned modeling methods by providing a different perspective on describing distributed systems and their interaction with context. It is suitable for building models that emphases on system structure and the interactions among the distinct component that constitute each system and its context. This methodology maps the systems structure and describes in details how each software component actually functions and how it interacts with other components and with context.
In other words, instead of focusing on the questions "What does the system do?" and "What is the reason that it operates on this way?" let us focus on "Where each operation is computed?" and "How does it work?". This is achieved by modeling each system as a network of Networked Turing machines where each machine simulates one distinct procedure of the system.
Especially on context-awareness, the benefit of using this approach is the development of a detailed description of the interaction between the system components and the context. As each machine simulates an algorithmic procedure built in the system, the produced model describes the way that context-awareness is achieved; not just the input and output entities of the procedures but also the details of how the application interacts with context.

## 5.2 Model formalism

The methodology produces models that follow the symbology and structure that is next presented.

*Symbology*

| | |
|---|---|
| : | binary operator of equivalence |
| // | comments or short descriptions |
| →, ↔, ← | data flow |

*Structure*
***abstract***. Text in plain language, it provides a description of the functionality of the system.

***procedures***. List and descriptions of the procedures executed in the system. Each one carries specific tasks and is represented in the model by one Networked Turing machine; it has an alphanumeric label in square brackets. In the graphs of the model each procedure is denoted by one node that has a numeric label e.g.
1 : [*input*] // user input
2 : [*screen_display*] // screen display
In this section designers may also include documentation about the algorithmic operations executed in each procedure. Documentation may include UML diagrams, pseudo code or any other document that efficiently describes the procedures.

***context***. Context entities and external sources. Each entity and source may output vectors of $C_A$ in one or more procedures. Also it may accept the output of procedures and change some the valuation in some vectors of $C_A$. Context labels are placed in parentheses. Their corresponding nodes in the graphs are labeled with letters e.g.
a : (*location*) // user or device location

***connections***. Connections among system procedures and between procedures and external sources. Connections in the model denoted like in the following examples.
*con*(*input*, *screen_display*) : [*input*] → [*screen_display*]
*con*(*location*, *GPS_unit*) : [*GPS_unit*] ← (*location*)
*con*(*server*, *client*) ∧ *con*(*client*, *server*) : [*server*] ↔ [*client*]

***graphs***. One or more graphs that represent procedures, context and connections.

# 6. A case study

Let us present two models for web application "WMS Map Viewer" (Rodis, 2017). The first model describes the application abstractly while the second provides a more refined description of the application. On this way we demonstrate how scalable models may be produced.

Model 1

***abstract***. “WMS Map Viewer” is a web application served by a cloud infrastructure and executed in any client device displaying web maps to its screen. Its functionality is to display interactive web maps on client devices. The application detects user interaction, user location and screen parameters. Then the app downloads the web maps requested by the user from the web services that provide them. Upon download, the application adjusts the maps on client device screen and displays them. Also the maps may be centered on user’s location.

***procedures***.

1 : [*soft_serve*] // Software served by cloud infrastructure

9 : [*client_app*] // The application running on client

***context***.

a : (*user*) // user of the client device

b : (*location*) // location unit (GPS) of client device

c : (*screen*) // screen of client device

d : (*providers*) // web map providers

***connections***.

[*soft_serve*] ↔ [*client_app*]

[*client_app*] ← (*user*) // detects user interaction

[*client_app*] ← (*location*) // retrieves location

[*client_app*] ↔ (*screen*)

// detects screen parameters and displays map content

[*client_app*] ↔ (*providers*)

// request and then downloads web maps from the providers

***graph***.

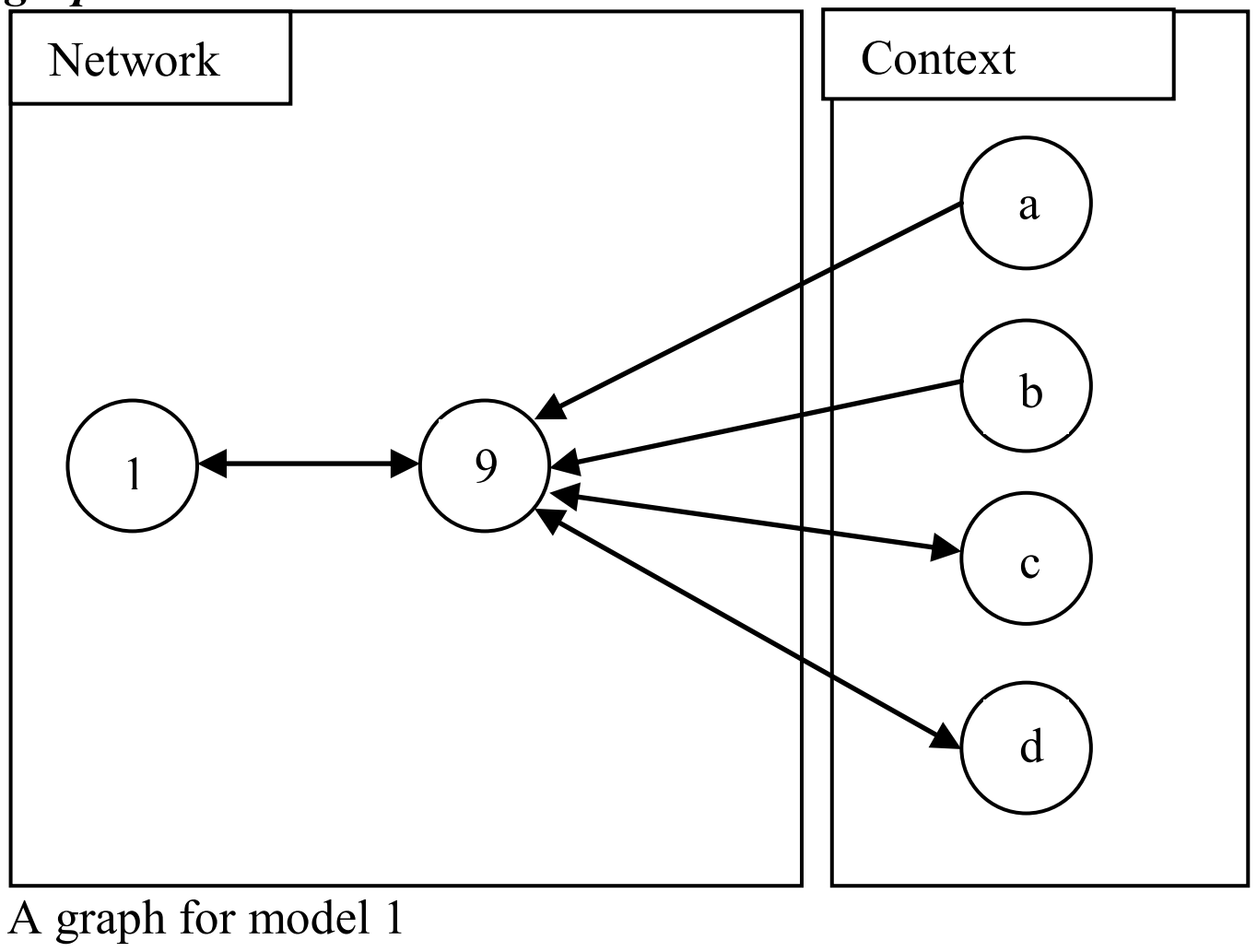

A graph for model 1

Model 2

***abstract***. “WMS Map Viewer” is an http / JavaScript web application for viewing and indexing web maps. The application is provided as SaaS (Software as a Service) served by a cloud infrastructure and it is designed so as to provide ubiquitous access to web maps for its users. In this case, the term “ubiquitous access” is used in the sense that the users of the application may access the content of web maps from any web enabled device; whether it is an old mobile phone and the users are on the field or a high-end system and the users are on their offices or labs. In all of these occasions the users will have the same quality of service using the application and the application will provide the same functionality through a User Interface (UI) that adapts the displayed content to the screen of each device.

The cloud infrastructure serves only the appropriate software components depending on the client device and the user requested functionality. Serving more software components than those that are necessary would increase network traffic and slow down the operation of low resources devices without any benefit. Also, the application detects user interaction so as to provide the desired content to the user and also detects the location of the user, after acquiring permission, so as to display the map content of the user’s location.

***procedures***.

1 : [*soft_serve*]

// software served by cloud infrastructure

2 : [*soft_download*]

// the application running on client downloads the necessary software components

3 : [*param_detection*]

// a procedure running on client detects device parameters asks for necessary UI

4 : [*map_display*]

// map content is adjusted and displayed on screen

5 : [*get_location*]

// location data are retrieved from device

6 : [*user_detection*]

// user interaction is detected

7 : [*map_request*]

// web maps are requested from external sources

8 : [*get_map*]

// web map content is downloaded from external sources

***context***.

a : (*user*) // user of the client device

b : (*location*) // location unit (GPS) of the device

c : (*screen*) // screen of client device

d : (*providers*) // web map providers

***connections***.

[*soft_serve*] ↔ [*soft_download*]

// cloud infrastructures accepts as input the request for serving software components and then uploads the necessary software on the client.

(*screen*) → [*param_detection*]

// a procedure running on client device accepts as input the screen parameters of the client device.

[*param_detection*] → [*soft_download*]

// based on screen parameters the necessary software is downloaded from cloud

[*soft_download*] → [*map_display*]

// the necessary software components are downloaded on the device so as to properly display map content

[*map_display*] → (*screen*)
// map content is adjusted and displayed on client screen
(*user*) → [*user_detection*]
// procedure [*user_detection*] accepts the variables of $C_A$ that describe user interaction
(*location*) → [*get_location*]
// the application retrieves GPS data.
[*get_location*] → [*map_request*]
[*user_detection*]→ [*map_request*]
// based on user interaction and user's location the client forms requests for web maps from map providers.
[*map_request*] → (*providers*)
// web maps are requested form map providers.
(*providers*) → [*get_map*]
// web maps are downloaded from external sources.
[*get_map*] → [*map_display*]
// downloaded web maps are inputted on map display procedure.

***graph***.

A graph that models the operation of WMS Map Viewer in details, model 2.

## 7. Discussion

The model of Networked Turing Machines extends current computational models by enclosing networking and interaction capabilities in the computational units. Using this model it is easy to describe distributed systems and systems of parallel processing as well as their interaction with external sources and other systems. These capabilities make this model ideal for describing context-aware applications and map their internal structure.

Computational complexity for context-awareness may also be studied using the above models. For instance, if a procedure requires exponential time to be completed then this affects the performance of the whole system. In particular, if the procedure of detecting some parameter of context corresponds to solving some worst case scenario of a problem in the NP complexity class, then the output of the system to the users or to other systems may not be computed efficiently.

The definitions of context and context-awareness presented in this paper are based on the theory of computation rather than a description of these notions in plain language as it usually happens on the literature of this field. On this way the above definitions are built on solid ground avoiding vagueness that may arise from unclear or disputed descriptions. The modeling methodology and the case study presented in the previous sections reveals the robustness of the models and definitions of this paper and provide a detailed map of the functionality of the system under study.